\documentclass[11pt]{article}
\usepackage{cite,emp,latexsym,amsmath,graphicx}
\makeatletter
\def\@normalsize{\@setsize\normalsize{12pt}\xpt\@xpt
\abovedisplayskip 10pt plus2pt minus5pt\belowdisplayskip \abovedisplayskip
\abovedisplayshortskip \z@ plus3pt\belowdisplayshortskip 6pt plus3pt
minus3pt\let\@listi\@listI}

\def\subsize{\@setsize\subsize{12pt}\xipt\@xipt}
\newcommand{\excise}[1]{}
\newcommand{\nix}[1]{}

\newcommand{\ket}[1]{|#1\rangle}
\newcommand{\proof}{\noindent\textit{Proof. }} 

\newcommand{\qed}{\mbox{$\Box$}}

\newcommand{\diag}{\textup{diag}}

\newcommand{\mat}[4]{\left(\begin{array}{cc}#1 & #2\\ #3 & #4\end{array}\right)}
\newcommand{\rmat}[4]{\left(\begin{array}{rr}#1 & #2\\ #3 & #4\end{array}\right)}

\newtheorem{theorem}{Theorem}
\newtheorem{lemma}[theorem]{Lemma}
\newtheorem{corollary}[theorem]{Corollary}

\newcommand{\C}{\mathbf{C}}
\newcommand{\R}{\mathbf{R}}

\makeatother

\begin{document}
\date{}

\title{\vspace{-1cm}\Large\textbf{The Simplified Toffoli Gate Implementation by Margolus is Optimal}} 

\author{
  Guang Song\footnote{
Guang Song's current address is: 
Baker Center for Bioinformatics and Biological Statistics, 
Iowa State University, 
Ames, IA 50011-3020, USA. Email: gsong@iastate.edu.} ~and Andreas Klappenecker\\
Department of Computer Science, Texas A\&M University,\\ 
College Station, TX~77843-3112, USA\\
 \{\texttt{gsong,klappi}\}\texttt{@cs.tamu.edu}
}

\maketitle

\begin{abstract}
\noindent 
Unitary operations are expressed in the quantum circuit model as a
finite sequence of elementary gates, such as controlled-not gates and
single qubit gates. We prove that the simplified Toffoli gate by
Margolus, which coincides with the Toffoli gate up to a single change
of sign, cannot be realized with less than three controlled-not
gates. If the circuit is implemented with three controlled-not gates,
then at least four additional single qubit gates are necessary.  This
proves that the implementation suggested by Margolus is optimal.
\end{abstract}

\def\thefootnote{\arabic{footnote}}
\setcounter{footnote}{0}

\section{Introduction}\label{sec:intro}\enlargethispage{4mm}
The simplified Toffoli gate realizes the unitary map $M\colon\C^8\rightarrow
\C^8$ given by 
\begin{eqnarray*}
\ket{00}\otimes \ket{\phi}  &\mapsto&  \ket{00}\otimes \phantom{Z}\ket{\phi}, \\
\ket{01}\otimes \ket{\phi}  &\mapsto&  \ket{01}\otimes \phantom{Z}\ket{\phi},	\\
\ket{10}\otimes \ket{\phi}  &\mapsto&  \ket{10}\otimes  Z \ket{\phi},	\\
\ket{11}\otimes \ket{\phi}  &\mapsto&  \ket{11}\otimes  X \ket{\phi},
\end{eqnarray*}
where $\ket{\phi}$ is an arbitrary state in $\C^2$, $X$ denotes the not gate, and $Z$ a phase gate, 
\begin{equation}\label{eq:gates}
 X = \mat{0}{1}{1}{0}, \qquad Z = \rmat{1}{0}{0}{-1}.
\end{equation}
The unitary map $M$ coincides with the Toffoli gate~\cite{bbc-egqc-95} on all vectors
of the standard basis, except that it maps the state $\ket{101}$ to
$-\ket{101}$ instead of $\ket{101}$; this strong resemblance explains
the name.

The simplified Toffoli gate has an elegant
implementation~\cite{bbc-egqc-95}, which is due to
Margolus~\cite{margolus94,d-qgc-98}, see
also~\cite{bbc-egqc-95,margolus02}. It merely requires three
controlled-not gates and four single qubit gates: \empprelude{input
qcg; prologues := 0;}
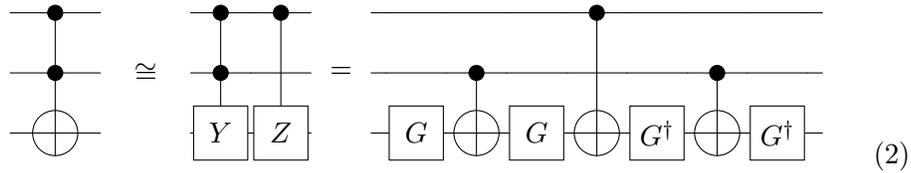
\begin{empfile}
\begin{equation}\label{circ:margolus}
\begin{emp}(50,50)
  setunit 2mm;
  qubits(3);
  
  cnot(icnd 1 , 2, gpos 0);
  label(btex $\cong$ etex,(QCxcoord+1/2QCstepsize, QCycoord[1]));
  QCxcoord := QCxcoord + QCstepsize;
  QCstepsize := QCstepsize-4mm;
  
  gate(icnd 2, 1, gpos 0, btex $Y$ etex);  
  gate(icnd 2, gpos 0, btex $Z$ etex);
  label(btex $=$ etex,(QCxcoord+1/2QCstepsize, QCycoord[1]));
  QCxcoord := QCxcoord + QCstepsize;

  wires(2mm);
  gate(gpos 0, btex $G$ etex);
  cnot(icnd 1, gpos 0);
  gate(gpos 0, btex $G$ etex);
  cnot(icnd 2, gpos 0);
  gate(gpos 0, btex $G^\dagger$ etex);
  cnot(icnd 1, gpos 0);
  gate(gpos 0, btex $G^\dagger$ etex);
  wires(2mm);
\end{emp}
\end{equation}
where
\begin{equation}\label{eq:gates2}
 Y=ZX=\rmat{0}{1}{-1}{0},\quad G = \rmat{\cos(\frac{1}{8}\pi)}{-\sin(\frac{1}{8}\pi)}
{\sin(\frac{1}{8}\pi)}{\cos(\frac{1}{8}\pi)}. 
\end{equation}
The congruence sign indicates equivalence up to a multiplication with
a diagonal matrix of phase factors. There are some alternatives to $M$
that differ in some other basis state by a sign, but that is not an 
essential change. 

The simplified Toffoli gate cannot substitute for the Toffoli gate in
general, because phase factors are important in true quantum
algorithms. However, if the Toffoli gates appear in pairs, then it is
possible to adapt the circuit structure to take advantage of the
simplified Toffoli gate~\cite{bbc-egqc-95}. The saving are quite
substantial in this case, because the best implementations of the
Toffoli gate known to date need \textit{fourteen} controlled-not and
single qubit gates.

Our main result expresses our appreciation of the beautiful structure
of the quantum circuit~(\ref{circ:margolus}). We prove that the
circuit is optimal in the following sense:
\par
\noindent\textbf{Theorem~M}
\textit{
Suppose that the simplified Toffoli gate $M$ is realized by a
sequence of controlled-not and single qubit gates. Any such sequence
contains at least three controlled-not gates. If it contains three
controlled-not gates, then at least four single qubit gates are
needed.}  
\medskip

Our proof reveals that the elegant structure of the
circuit~(\ref{circ:margolus}) is not an arbitrary artifact. Indeed,
any optimal quantum circuit realizing $M$ is essentially of this form,
except that the single qubit gates are possibly different.

We followed the seminal paper~\cite{bbc-egqc-95} in our choice of the
universal set of gates, because this has been adopted in several
textbooks as well. Choosing controlled-not gates and single qubit
gates is somewhat arbitrary, but similar arguments can be carried out
for other universal sets of quantum gates. The main reason for our
choice is that the number of controlled-not and single qubit gates
constitute the prevailing measure of complexity currently used in Computer
Science. This paper is part of a larger
program, where we try to gain a better understanding of basic quantum
circuit structures.

\medskip
\noindent\textit{Notations.} 
In addition to the gates introduced in (\ref{eq:gates}) and (\ref{eq:gates2}), we use 
$$ 
H = \frac{1}{\sqrt{2}}\rmat{1}{1}{1}{-1}
$$ 
to denote the Hadamard gate. The state $\ket{0_x}=H\ket{0}$ 
denotes the eigenstate~of~$X$ with eigenvalue~$+1$. 
The most significant qubit is represented by the topmost wire in the quantum
circuit notation, and the least significant qubit by the lowest wire. 
We denote by $\C$ the field of complex numbers, and
by $\R$ the field of real numbers.

\section{Review of Previous Work}\label{sec:previous}
We review in this section some useful lemmas, all of which are proved 
in~\cite{SongKlappi03}. We will use these results in the proof of Theorem~M. 
Recall that it is possible to switch the control and the target
qubit of a controlled-not gate by conjugation with Hadamard matrices
$H$:
\begin{equation}\label{circ:flip}
\begin{emp}(50,50)
  setunit 2mm; qubits(2);
                                                                                
  cnot(icnd 0, gpos 1);
  label(btex $=$ etex,(QCxcoord+1/2QCstepsize, QCycoord[0]+3mm));
  QCxcoord := QCxcoord + QCstepsize;
  QCstepsize := QCstepsize-4mm;
  wires(2mm);
  gate(gpos 1, btex $H$ etex, 0, btex $H$ etex);
  cnot(icnd 1, gpos 0);
  gate(gpos 1, btex $H$ etex, 0, btex $H$ etex);
  wires(2mm);
\end{emp}
\end{equation}
Due to this important fact, it suffices to consider controlled-not
gates where the control is on a higher significant qubit that the target qubit
when we write down the general form of a circuit.  

\begin{lemma}\label{lemma:entanglement}
Let $\ket{\psi}$, $\ket{\phi}$ be nonzero elements of $\C^2$.
The input\/
$\ket{\psi}\otimes \ket{\phi}$ to a controlled-$U$ gate will produce
an entangled output state if and only if\/ $\ket{\phi}$ is not an eigenvector of\/ $U$ and\/ $\ket{\psi}=a\ket{0}+b\ket{1}$
with\/ $a,b\neq 0$.
\end{lemma}

\begin{lemma}\label{lemma:diag}
Assume that $\ket{\phi}$ is an eigenvector of a unitary $2\times 2$
matrix~$U$ with eigenvalue~$\lambda_\phi$. Let $\ket{\psi}$ denote a
state in $\C^2$.  If we input $\ket{\psi}\otimes \ket{\phi}$ to the
controlled-$U$ gate, then the output is of the form
$\diag(1,\lambda_\phi)\ket{\psi}\otimes \ket{\phi}$.  In particular,
the output is not entangled.
\end{lemma}

A controlled-$U$ gate can be realized with two controlled-not gates and 
several single qubit gates, as follows: 
\begin{equation}\label{circ:double}
\vbox{\hbox{
\begin{emp}(50,50)
  setunit 2mm;
  qubits(2);
                                                                                
  gate(icnd 1, gpos 0, btex $U$ etex);
  label(btex $=$ etex,(QCxcoord+1/2QCstepsize, QCycoord[0]+3mm));
  QCxcoord := QCxcoord + QCstepsize;
  QCstepsize := QCstepsize-4mm;
  wires(2mm);
  gate(gpos 1, btex $A_1$ etex, 0, btex $B_1$ etex);
  cnot(icnd 1, gpos 0);
  gate(gpos 1, btex $A_2$ etex, 0, btex $B_2$ etex);
  cnot(icnd 1, gpos 0);
  gate(gpos 1, btex $A_3$ etex, 0, btex $B_3$ etex);
                                                                                
  wires(2mm);
\end{emp}
}}
\end{equation}
The following two lemmas describe some constraints on the gates $A_1,
A_2$, and $A_3$. We call a single qubit gate sparse if it is realized
by a diagonal or antidiagonal $2\times 2$ matrix. 

\begin{lemma}\label{lemma:sparse}
If the matrix $A_1$ is sparse,
then $A_2, A_3$  are sparse as well.
\end{lemma}

\begin{lemma}\label{lemma:notsparse}
Suppose that $U$ is not a multiple of the identity matrix.
If $A_1$ in the circuit (\ref{circ:double}) is
not sparse, then $A_2, A_3$ are not sparse either.
\end{lemma}

\section{Proof of Theorem~M}
We proceed to show that three controlled-not gates are necessary and
sufficient in any realization of unitary map $M$ by a sequence of
controlled-not and single qubit gates. We first prove that at least
two controlled-not operations act on the last qubit:
\begin{lemma}\label{lemma:oneonlast}
Suppose there are some interactions between the top two qubits, but
only one controlled-not interaction between the two control qubits 
and the target bit. The circuit cannot realize the Margolus map $M$.
\end{lemma}

\proof Any such circuit can be represented in the form
\begin{equation}
\begin{emp}(50,50)
  setunit 2mm;
    qubits(3);
    circuit(1cm)(gpos 1,2, btex $B_1$ etex);
    gate(gpos 0, btex $C_1$ etex);
    cnot(icnd 1, gpos 0)
    gate(gpos 0, btex $C_2$ etex);
    circuit(1cm)(gpos 1,2, btex $B_2$ etex);
\end{emp}
\end{equation}
Let $\ket{0_x}=\frac{1}{\sqrt{2}}(\ket{0}+\ket{1})$, and denote by
$\ket{\varphi}$ an arbitrary state in $\C^4$. If we input
$\ket{\varphi}\otimes C_1^\dagger\ket{0_x}$ to the above quantum
circuit, then the least significant qubit of the output state is not
entangled with the remaining two qubits, regardless of the nature of
input state $\ket{\varphi}$.

However, if we choose the input 
$\ket{00}\otimes \ket{\phi} + \ket{10}\otimes \ket{\phi} + \ket{11}\otimes \ket{\phi}$,
then the output of $M$ is 
$\ket{00}\otimes \ket{\phi} + \ket{10}\otimes Z\ket{\phi} + \ket{11}\otimes X\ket{\phi}$. Note that the target qubit is entangled with the other two qubits, since $\ket{\phi}$ cannot be an eigenvector
of $X$ and $Z$ at the same time. Contradiction.~\qed

\begin{corollary}\label{corollary}
The target qubit is affected by at least two controlled-not operations. 
\end{corollary}

\paragraph{Two Controlled-Not Gates.}
Assume there are only two controlled-not gates in the circuit. Taking
Corollary~\ref{corollary} and the identity~(\ref{circ:flip}) into
account, we may assume that both controlled-not gates operate on the
target bit. The control qubits of the two gates have to be different,
for otherwise it would not be possible to entangle all input qubits
with the outout qubit.

Since $M=M^\dagger$, we do not need to concern ourselves with the
order the two controlled-not gate in such a circuit. Therefore, we may assume
that the circuit is of the form 
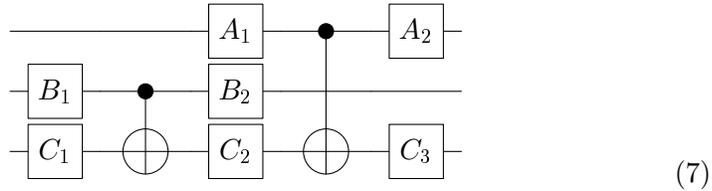
\begin{equation}\label{circ:twocnot}
\begin{emp}(50,50)
  setunit 2mm;
      qubits(3);
      gate(gpos 1, btex $B_1$ etex, 0, btex $C_1$ etex);
      cnot(icnd 1, gpos 0)
      gate(gpos 1, btex $B_2$ etex, 2, btex $A_1$ etex, 0, btex $C_2$ etex);
      cnot(icnd 2, gpos 0)
      gate(gpos 2, btex $A_2$ etex,0, btex $C_3$ etex);
\end{emp}
\end{equation}

\begin{lemma}\label{lemma:twocnot}
The circuit~(\ref{circ:twocnot}) cannot implement the simplified 
Toffoli gate.
\end{lemma}
\proof When the top qubit is $\ket{0}$, then the circuit~(\ref{circ:twocnot})
can still entangle the least significant two qubits, 
contradicting the behavior of $M$.~\qed.

\begin{corollary}\label{corollary2}
At least three controlled-not gates are necessary in an implementation
of the simplified Toffoli gate $M$ by a sequence of controlled-not and
single qubit gates.
\end{corollary}

\paragraph{Three Controlled-Not Gates.}
The remaining argument proceeds by considering all possible
configurations of the three controlled-not gates. Initially, we allow
an arbitrary number of single qubit operations. Thus, we may assume
that the target qubit has lesser significance than the control qubit
by applying~(\ref{circ:flip}), so that we have to consider ${ 3
\choose 2}^3=27$ configurations of controlled-not gates.  We use the pictogram
\begin{center}
\begin{emp}(50,50)
  def cnotconf(expr num) = 
     begingroup
       if num=0: cnot(icnd 1, gpos 0); fi;
       if num=1: cnot(icnd 2, gpos 0); fi;
       if num=2: cnot(icnd 2, gpos 1); fi;
     endgroup; 
   enddef;
   def conff(expr a,b,c) = 
     begingroup 
       cnotconf(a); 
       cnotconf(b); 
       cnotconf(c);
       QCxcoord := QCxcoord + 5mm; 
     endgroup;
   enddef;
  setunit 0.5mm;
  qubits(3);
  conff(0,2,1);
\end{emp}
\end{center}
as a shorthand for a general quantum circuit of the
form~(\ref{circ:2diffcontrol}) that contains in addition to the specified controlled-not configuration all potential single qubit gates. 
We distinguish three different cases
that we record here for the orientation of the reader:
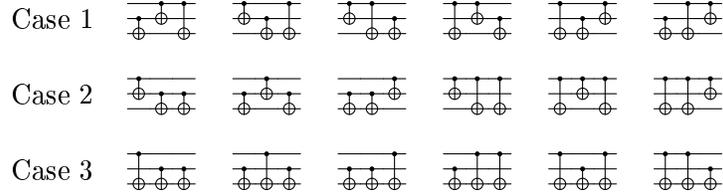
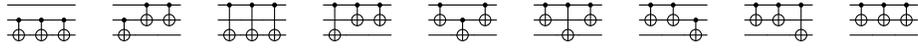
\begin{center}
\begin{emp}(50,50)
   def cnotconf(expr num) = 
     begingroup
       if num=0: cnot(icnd 1, gpos 0); fi;
       if num=1: cnot(icnd 2, gpos 0); fi;
       if num=2: cnot(icnd 2, gpos 1); fi;
     endgroup; 
   enddef;
   def conff(expr a,b,c) = 
     begingroup 
       cnotconf(a); 
       cnotconf(b); 
       cnotconf(c);
       QCxcoord := QCxcoord + 5mm; 
     endgroup;
   enddef;
   setunit 0.5mm;
   qubits(3);
   numeric basex;
   numeric basey;
   basex := QCxcoord;
   basey := QCycoord[0];

   label(btex Case 1 etex, (QCxcoord,QCycoord[1]));
   QCxcoord := QCxcoord + 10mm;
  conff(0,2,1);
  conff(2,0,1);
  conff(2,1,0);
  conff(1,2,0);
  conff(1,0,2);
  conff(0,1,2);

   basey := basey - 10mm;
   QCxcoord := basex;
   QCyofs := basey;
   initycoord;

   label(btex Case 2 etex, (QCxcoord,QCycoord[1]));
   QCxcoord := QCxcoord + 10mm;
   conff(2,0,0);
   conff(0,2,0);
   conff(0,0,2);
   conff(2,1,1);
   conff(1,2,1);
   conff(1,1,2);

 basey := basey - 10mm;
   QCxcoord := basex;
   QCyofs := basey;
   initycoord;

   label(btex Case 3 etex, (QCxcoord,QCycoord[1]));
   QCxcoord := QCxcoord + 10mm;
   conff(1,0,0);
   conff(0,1,0);
   conff(0,0,1);
   conff(0,1,1);
   conff(1,0,1);
   conff(1,1,0);     
\end{emp}
\end{center}
The remaining configurations 
\begin{center}
\begin{emp}(50,50)
  setunit 0.5mm;
  qubits(3);
  conff(0,0,0);
  conff(0,2,2);
  conff(1,1,1);
  conff(1,2,2);
  conff(2,0,2); 
  conff(2,1,2);
  conff(2,2,0);
  conff(2,2,1);
  conff(2,2,2);
\end{emp}
\end{center}
are excluded because they are ruled out by 
Corollary~\ref{corollary} or lack the capability to
entangle all three qubits, hence cannot implement $M$.

It turns out that only a single configuration allows to realize the
simplified Toffoli gate~$M$. In most cases, we are able to exclude
circuit structures because they exhibit entanglement properties that
are inconsistent with the behavior of the simplified Toffoli
gate~$M$. We record the following trivial observation:
\begin{lemma}\label{lemma:remainentangled}
Suppose that two systems $A$ and $B$ of qubits are entangled. If the
remaining gate operations affect the systems $A$ and $B$ separately,
then $A$ and $B$ remain entangled. 
\end{lemma}
\proof Seeking a contradiction, we assume that the resulting output
state is not entangled, i.e., is of the form $\ket{\phi_A}\otimes
\ket{\phi_B}$. The gate operations leading to this output state can be
written in the form $U_A\otimes U_B$, since the operations affect the
systems separately. This would imply that the input state $U_A^\dagger
\ket{\phi_A}\otimes U_B^\dagger \ket{\phi_B}$ was separable as well, 
in contradiction to the assumption.~\qed

\medskip
\noindent\textit{Case 1.}  The circuits configuration are in this case
characterized by the fact that exactly two controlled-not gates act on
the target qubit of~$M$, and they are controlled from different qubits.
The third controlled-not gate is between the two most
significant qubits.  It turns out that none of these circuits can
implement $M$, even if we allow single qubit gates on all possible
positions.
\begin{lemma}\label{lemma:2diffcontrol}
A circuit with one of the configurations
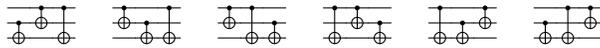
\begin{center}
\begin{emp}(50,50)
  setunit 0.5mm;
  qubits(3);
  conff(0,2,1);
  conff(2,0,1);
  conff(2,1,0);
  conff(1,2,0);
  conff(1,0,2);
  conff(0,1,2);
\end{emp}
\end{center}
cannot implement the simplified Toffoli gate~$M$.
\end{lemma}
\proof
The most general circuit corresponding to the first pictogram is given by
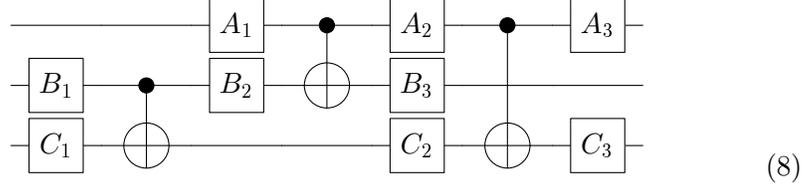
\begin{equation}\label{circ:2diffcontrol}
\begin{emp}(50,50)
  setunit 2mm;
      qubits(3);
      gate(gpos 1, btex $B_1$ etex, 0, btex $C_1$ etex);
      cnot(icnd 1, gpos 0)
      gate(gpos 1, btex $B_2$ etex, 2, btex $A_1$ etex);
      cnot(icnd 2, gpos 1)
      gate(gpos 2, btex $A_2$ etex, 1, btex $B_3$ etex, 0, btex $C_2$ etex);
      cnot(icnd 2, gpos 0)
      gate(gpos 0, btex $C_3$ etex, 2, btex $A_3$ etex);
\end{emp}
\end{equation}
Seeking a contradiction, we assume that $A_1$ is not sparse.  If we
provide an input $\ket{0}\otimes \ket{\phi}\otimes
C_1^{\dagger}\ket{0_x}$,
then the top two quantum bits could possibly get
entangled by Lemma~\ref{lemma:entanglement}, and if so, these two qubits would 
remain entangled by Lemma~\ref{lemma:remainentangled}, contradicting the
behavior of $M$. Hence, $A_1$ must be sparse. 

By the same token, $A_2$ has to be sparse, because otherwise we can find an input state of the form $\ket{0}\otimes \ket{\varphi}$, such that the most
and least significant qubit get entangled, contradicting the behavior
of $M$. As a consequence, $A_3$ is sparse as well.

Therefore, the behavior of the circuit~(\ref{circ:2diffcontrol}) on input of $\ket{0}\otimes \ket{\varphi}$, $\ket{\varphi}\in \C^4$, can be simulated by a circuit of the form
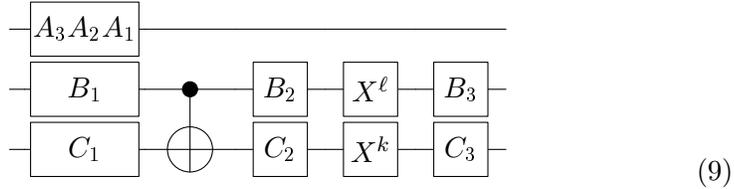
\begin{equation}\label{circ:2diffcontrol2}
\begin{emp}(50,50)
  setunit 2mm;
      qubits(3);
      wires(2mm);
      widegate(2);
      gate(gpos 2, btex $A_3A_2A_1$ etex, 
                1, btex $B_1$ etex, 
                0, btex $C_1$ etex);
      normalgate;
      QCgatewidth := 1;
      QCstepsize := 6*unit;
      cnot(icnd 1, gpos 0)
      gate(gpos 1, btex $B_2$ etex, 0, btex $C_2$ etex);
      gate(gpos 1, btex $X^\ell$ etex, 0, btex $X^k$ etex);
      gate(gpos 1, btex $B_3$ etex, 0, btex $C_3$ etex);
\end{emp}
\end{equation}
where the values of $k,\ell\in \{0,1\}$ depend on whether $A_1$ and $A_2$ are diagonal or antidiagonal. It is obvious from this circuit that we can choose a separable state
$\ket{\varphi}$ such that the two least significant qubits get entangled,
even though the topmost qubit is in the state $\ket{0}$. This
contradicts the behavior of $M$, thus a circuit of the form
(\ref{circ:2diffcontrol}) cannot realize $M$.

In the same way, it is straightforward to see that 
neither the second nor the third pictogram can realize $M$.

Finally, the last three pictograms represent 
the inverse circuits of the first three, so none of them can realize 
$M$ since $M$ is self-inverse.~\qed

\medskip
\noindent\textit{Case 2.}  The circuit configurations of the second
case are characterized by the fact that exactly two controlled-not
gates act on the target qubit, and both are controlled from the same
qubit. We distinguish in our discussion whether they are controlled by
the middle qubit (\textit{Case 2.1}), or by the most significant qubit
(\textit{Case 2.2}).
\medskip

\noindent\textit{Case 2.1.}  This case treats the configurations of controlled-not gates that have the pictorial representation
\begin{center}
\begin{emp}(50,50)
  setunit 0.5mm;
  qubits(3);
  conff(2,0,0);
  conff(0,0,2);
  conff(0,2,0);
\end{emp}
\end{center}

\begin{lemma}\label{lemma:a1sparse}
A circuit with one of the controlled-not configurations
\begin{center}
\begin{emp}(50,50)
  setunit 0.5mm;
  qubits(3);
  conff(2,0,0);
  conff(0,0,2);
\end{emp}
\end{center}
cannot implement the simplified Toffoli gate~$M$.
\end{lemma}
\proof 
The most general circuit corresponding to the first pictogram is given by 
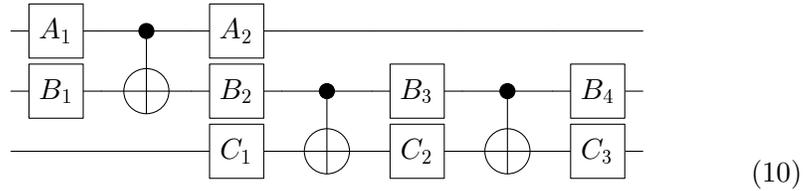
\begin{equation}\label{circ:2sameCntrl1a}
\begin{emp}(50,50)
  setunit 2mm;
      qubits(3);
      gate(gpos 2, btex $A_1$ etex, 1, btex $B_1$ etex);
      cnot(icnd 2, gpos 1)
      gate(gpos 2, btex $A_2$ etex, 1, btex $B_2$ etex, 0, btex $C_1$ etex);
      cnot(icnd 1, gpos 0)
      gate(gpos 1, btex $B_3$ etex, 0, btex $C_2$ etex);
      cnot(icnd 1, gpos 0)
      gate(gpos 1, btex $B_4$ etex, 0, btex $C_3$ etex);
\end{emp}
\end{equation}
The second pictogram is covered by taking the inverse of the above
circuit, hence does not need to be treated separately. If we input
$\ket{0}\otimes B_1^\dagger \ket{0}\otimes \ket{0}$, then the circuit
will produce an entangled output state by
Lemmas~\ref{lemma:entanglement} and~\ref{lemma:remainentangled}
if $A_1$ is not sparse. Hence, $A_1$ has to be sparse, and
consequently $A_2$ as well.

If we take the input state $\ket{0}\otimes \ket{\varphi},$ with $\ket{\varphi}\in \C^4$, then 
the circuit~(\ref{circ:2sameCntrl1a}) has to act identically on 
$\ket{\varphi}$. Consequently, we obtain the circuit identity
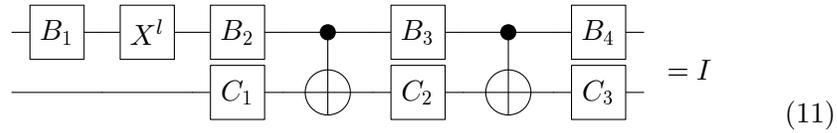
\begin{equation}\label{circ:2sameCntrl1a2}
\begin{emp}(50,50)
  setunit 2mm;
      qubits(2);
      gate(gpos 1, btex $B_1$ etex);
      gate(gpos 1, btex $X^l$ etex);
      gate(gpos 1, btex $B_2$ etex, 0, btex $C_1$ etex);
      cnot(icnd 1, gpos 0)
      gate(gpos 1, btex $B_3$ etex, 0, btex $C_2$ etex);
      cnot(icnd 1, gpos 0)
      gate(gpos 1, btex $B_4$ etex, 0, btex $C_3$ etex);
      label(btex $= I$ etex,(QCxcoord+1/2QCstepsize, QCycoord[0]+3mm));
      QCxcoord := QCxcoord + QCstepsize;
      QCstepsize := QCstepsize-4mm;
\end{emp}
\end{equation}
On the other hand, we can derive from the input $\ket{1}\otimes
\ket{\varphi}$ the circuit identity
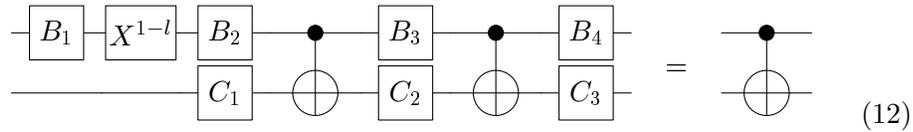
\begin{equation}\label{circ:2sameCntrl1a3}
\begin{emp}(50,50)
  setunit 2mm;
      qubits(2);
      gate(gpos 1, btex $B_1$ etex);
      widegate(1.3);
      gate(gpos 1, btex $X^{1-l}$ etex);
      normalgate;
      gate(gpos 1, btex $B_2$ etex, 0, btex $C_1$ etex);
      cnot(icnd 1, gpos 0)
      gate(gpos 1, btex $B_3$ etex, 0, btex $C_2$ etex);
      cnot(icnd 1, gpos 0)
      gate(gpos 1, btex $B_4$ etex, 0, btex $C_3$ etex);
      label(btex $=$ etex,(QCxcoord+1/2QCstepsize, QCycoord[0]+3mm));
      QCxcoord := QCxcoord + QCstepsize;

      cnot(icnd 1, gpos 0); 	
      
      
\end{emp}
\end{equation}
Combining the previous two circuit identities, we obtain 
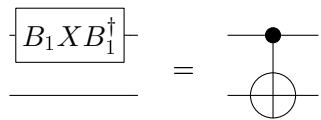
\begin{center}
\begin{emp}(50,50)

      setunit 2mm;
      qubits(2);
     
      widegate(2);
      gate(gpos 1, btex $B_1XB_1^{\dagger}$ etex);
      normalgate;
      wires(1mm);
 
      label(btex $=$ etex,(QCxcoord+1/2QCstepsize, QCycoord[0]+3mm));
      QCxcoord := QCxcoord + QCstepsize;

      cnot(icnd 1, gpos 0);
\end{emp}
\end{center}
which is absurd. Therefore, a circuit of the form~(\ref{circ:2sameCntrl1a}) 
cannot implement~$M$.~\qed

\begin{lemma}
A circuit with controlled-not configuration
\begin{center}
\begin{emp}(50,50)
  setunit 0.5mm;
  qubits(3);
  conff(0,2,0);
\end{emp}
\end{center}
cannot implement the simplified Toffoli gate~$M$.
\end{lemma}
\proof The most general circuit corresponding to the configuration depicted
in the last pictogram is given by 
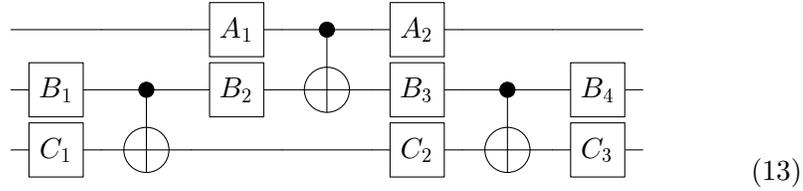
\begin{equation}\label{circ:2sameCntrl1b}
\begin{emp}(50,50)
  setunit 2mm;
      qubits(3);
      gate(gpos 1, btex $B_1$ etex, 0, btex $C_1$ etex);
      cnot(icnd 1, gpos 0)
      gate(gpos 2, btex $A_1$ etex, 1, btex $B_2$ etex);
      cnot(icnd 2, gpos 1)
      gate(gpos 2, btex $A_2$ etex, 1, btex $B_3$ etex, 0, btex $C_2$ etex);
      cnot(icnd 1, gpos 0)
      gate(gpos 1, btex $B_4$ etex, 0, btex $C_3$ etex);
\end{emp}
\end{equation}
Once again, we want to show that this circuit structure cannot
implement~$M$.  We note that $A_1$ and $A_2$ have to be sparse (for
otherwise it would be possible to find a state $\ket{\varphi}\in \C^4$
such that $\ket{0}\otimes \ket{\varphi}$ leads to an entangled output
state, arguing as in the previous case). The
circuit~(\ref{circ:2sameCntrl1b}) has to act as the identity an input
state $\ket{0}\otimes \ket{\varphi}$. Thus, we obtain the circuit
identity
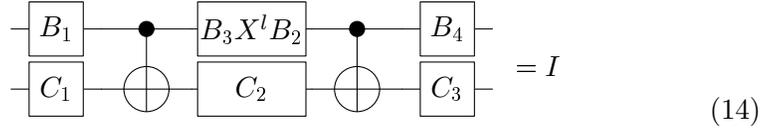
\begin{equation}\label{circ:2sameCntrl1b1}
\begin{emp}(50,50)
  setunit 2mm;
      qubits(2);
      gate(gpos 1, btex $B_1$ etex, 0, btex $C_1$ etex);
      cnot(icnd 1, gpos 0)
      widegate(2);
      gate(gpos 1, btex $B_3X^lB_2$ etex, 0, btex $C_2$ etex);
      normalgate;
      cnot(icnd 1, gpos 0)
      gate(gpos 1, btex $B_4$ etex, 0, btex $C_3$ etex);
      label(btex $= I$ etex,(QCxcoord+1/2QCstepsize, QCycoord[0]+3mm));
      QCxcoord := QCxcoord + QCstepsize;
      QCstepsize := QCstepsize-4mm;
\end{emp}
\end{equation}
Similarly, the input $\ket{1}\otimes \ket{\phi}\otimes \ket{\psi}$ leads to the circuit identity
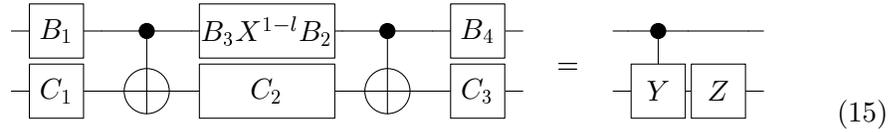
\begin{equation}\label{circ:2sameCntrl1b2}
\begin{emp}(50,50)
  setunit 2mm;
      qubits(2);
      gate(gpos 1, btex $B_1$ etex, 0, btex $C_1$ etex);
      cnot(icnd 1, gpos 0)
      widegate(2.5);
      gate(gpos 1, btex $B_3X^{1-l}B_2$ etex, 0, btex $C_2$ etex);
      normalgate;
      cnot(icnd 1, gpos 0)
      gate(gpos 1, btex $B_4$ etex, 0, btex $C_3$ etex);
      label(btex $= $ etex,(QCxcoord+1/2QCstepsize, QCycoord[0]+3mm));
      QCxcoord := QCxcoord + QCstepsize;
      QCstepsize := QCstepsize-4mm;
      
      wires(2mm);
      gate(icnd 1, gpos 0, btex $Y$ etex);
      gate(gpos 0, btex $Z$ etex);      
      wires(2mm);
\end{emp}
\end{equation}
From circuits~(\ref{circ:2sameCntrl1b1}) and (\ref{circ:2sameCntrl1b2}), we have,
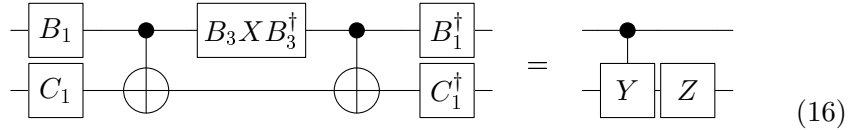
\begin{equation}\label{circ:2sameCntrl1b3}
\begin{emp}(50,50)
  setunit 2mm;
      qubits(2);
      gate(gpos 1, btex $B_1$ etex, 0, btex $C_1$ etex);
      cnot(icnd 1, gpos 0)
      widegate(2);
      gate(gpos 1, btex $B_3XB_3^\dagger$ etex);
      normalgate;
      cnot(icnd 1, gpos 0)
      gate(gpos 1, btex $B_1^{\dagger}$ etex, 0, btex $C_1^{\dagger}$ etex);
      label(btex $= $ etex,(QCxcoord+1/2QCstepsize, QCycoord[0]+3mm));
      QCxcoord := QCxcoord + QCstepsize;
      QCstepsize := QCstepsize-4mm;
      wires(2mm);
      gate(icnd 1, gpos 0, btex $Y$ etex);
      gate(gpos 0, btex $Z$ etex);
      wires(2mm);
\end{emp}
\end{equation}
Moving $Z$ to the left-hand side of the equation, we have
\begin{equation}\label{circ:2sameCntrl1b4}
\begin{emp}(50,50)
  setunit 2mm;
      qubits(2);
      wires(1mm);
      gate(gpos 1, btex $B_1$ etex, 0, btex $C_1$ etex);
      normalgate;
      cnot(icnd 1, gpos 0);
      widegate(2);
      gate(gpos 1, btex $B_3 XB_3^\dagger$ etex);
      normalgate;
      cnot(icnd 1, gpos 0)
      widegate(1.5);
      gate(gpos 1, btex $B_1^{\dagger}$ etex, 0, btex $ZC_1^{\dagger}$ etex);
      label(btex $= $ etex,(QCxcoord+1/2QCstepsize, QCycoord[0]+3mm));
      QCxcoord := QCxcoord + QCstepsize;
      QCstepsize := QCstepsize-4mm;
      wires(2mm);
      normalgate;
      gate(icnd 1, gpos 0, btex $Y$ etex);
      wires(2mm);
\end{emp}
\end{equation}
Now for any input state $\ket{\phi}\otimes
(C_1)^{\dagger}\ket{0_x}$, such that 
$\ket{\phi}=a\ket{0}+b\ket{1}$ is in superposition, $a,b\neq 0$, 
the circuit on the
left hand side does not entangle the two qubits. Therefore, by
Lemma~\ref{lemma:entanglement}, we know that $(C_1)^{\dagger}\ket{0_x}$
has to be an eigenvector of $Y$. 
Furthermore, by Lemma~\ref{lemma:diag}, we have
\begin{equation}
B_1^{\dagger}B_3^{\dagger}XB_3B_1 = \diag(1,y_0),
\end{equation}
where $y_0$ is one the eigenvalues of $Y$, i.e., $y_0 = (-i)$ or $i$.
Taking the trace on both sides, we get a contradiction.~\qed

\medskip

\noindent\textit{Case 2.2.}
We now treat the configurations that have the pictorial representation 
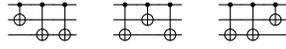
\begin{center}
\begin{emp}(50,50)
  setunit 0.5mm;
  qubits(3);
  conff(2,1,1);
  conff(1,2,1);
  conff(1,1,2);
\end{emp}
\end{center}
Changing the role of the two most significant qubits, we arrive at simplified Toffoli gate $M'$ that is implemented by 
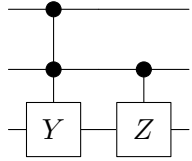
\begin{center}
\begin{emp}(50,50)
  setunit 2mm;
  qubits(3);

  gate(icnd 1,2, gpos 0, btex $Y$ etex);
  gate(icnd 1, gpos 0, btex $Z$ etex);
\end{emp}
\end{center}
Thus, it is equivalent to use the circuits~(\ref{circ:2sameCntrl1a}) and 
(\ref{circ:2sameCntrl1b}) of the previous case to implement $M'$.  

\begin{lemma}\label{lemma:mprime}
Circuits with configuration 
\begin{center}
\begin{emp}(50,50)
  setunit 0.5mm;
  qubits(3);
  conff(2,1,1);
  conff(1,1,2);
\end{emp}
\end{center}
cannot realize the Toffoli gate $M$. 
\end{lemma}
\proof We use the above trick and show that circuit~(\ref{circ:2sameCntrl1a}) cannot implement~$M'$. We derive from the input state 
$\ket{0}\otimes \ket{\phi}\otimes \ket{\psi}$ the circuit identity
\begin{equation}\label{circ:2sameCntrl2a1}
\begin{emp}(50,50)
  setunit 2mm;
      qubits(2);
      wires(1mm);
      widegate(2);
      gate(gpos 1, btex $B_2X^{l}B_1$ etex, 0, btex $C_1$ etex);
      normalgate;
      cnot(icnd 1, gpos 0)
      gate(gpos 1, btex $B_3$ etex, 0, btex $C_2$ etex);
      cnot(icnd 1, gpos 0)
      gate(gpos 1, btex $B_4$ etex, 0, btex $C_3$ etex);
      label(btex $= $ etex,(QCxcoord+1/2QCstepsize, QCycoord[0]+3mm));
      QCxcoord := QCxcoord + QCstepsize;
      QCstepsize := QCstepsize-4mm;
      wires(2mm);
      gate(icnd 1, gpos 0, btex $Z$ etex);
      wires(2mm);
\end{emp}
\end{equation}
Similarly, the input state $\ket{1}\otimes \ket{\phi}\otimes \ket{\psi}$ yields
\begin{equation}\label{circ:2sameCntrl2a2}
\begin{emp}(50,50)
  setunit 2mm;
      qubits(2);
      wires(1mm);
      widegate(2.5);
      gate(gpos 1, btex $B_2X^{1-l}B_1$ etex, 0, btex $C_1$ etex);
      normalgate;
      cnot(icnd 1, gpos 0)
      gate(gpos 1, btex $B_3$ etex, 0, btex $C_2$ etex);
      cnot(icnd 1, gpos 0)
      gate(gpos 1, btex $B_4$ etex, 0, btex $C_3$ etex);
      label(btex $= $ etex,(QCxcoord+1/2QCstepsize, QCycoord[0]+3mm));
      QCxcoord := QCxcoord + QCstepsize;
      QCstepsize := QCstepsize-4mm;
      wires(2mm);
      gate(icnd 1, gpos 0, btex $X$ etex);
      wires(2mm);
\end{emp}
\end{equation}
We can deduce from circuits~(\ref{circ:2sameCntrl2a1}) and (\ref{circ:2sameCntrl2a2}) the relation 
\begin{equation}\label{circ:2sameCntrl2a3}
\begin{emp}(50,50)
  setunit 2mm;
  qubits(2);
  wires(1mm);
  widegate(2);
  gate(gpos 1, btex $B_1^{\dagger}XB_1$ etex);
  normalgate;
  gate(icnd 1, gpos 0,  btex $Z$ etex);
  label(btex $= $ etex,(QCxcoord+1/2QCstepsize, QCycoord[0]+3mm));
  QCxcoord := QCxcoord + QCstepsize;
  QCstepsize := QCstepsize-4mm;
  wires(2mm);
  gate(icnd 1, gpos 0, btex $X$ etex);
  wires(2mm);
\end{emp}
\end{equation}
which clearly leads to a contradiction.~\qed

\begin{lemma}
Circuits with configuration 
\begin{center}
\begin{emp}(50,50)
  setunit 0.5mm;
  qubits(3);
  conff(1,2,1);
\end{emp}
\end{center}
cannot realize the Toffoli gate $M$. 
\end{lemma}
\proof It suffices to show that circuit~(\ref{circ:2sameCntrl1b})
cannot implement $M'$. Considering the input state $\ket{0}\otimes \ket{\phi}\otimes \ket{\psi}$, we obtain the circuit identity
\begin{equation}\label{circ:2sameCntrl2b1}
\begin{emp}(50,50)
  setunit 2mm;
      qubits(2);
      gate(gpos 1, btex $B_1$ etex, 0, btex $C_1$ etex);
      cnot(icnd 1, gpos 0);
      widegate(2);
      gate(gpos 1, btex $B_3X^lB_2$ etex, 0, btex $C_2$ etex);
      normalgate;
      cnot(icnd 1, gpos 0)
      gate(gpos 1, btex $B_4$ etex, 0, btex $C_3$ etex);
      label(btex $= $ etex,(QCxcoord+1/2QCstepsize, QCycoord[0]+3mm));
      QCxcoord := QCxcoord + QCstepsize;
      QCstepsize := QCstepsize-4mm;
      wires(2mm);
      gate(icnd 1, gpos 0, btex $Z$ etex);
      wires(2mm);
\end{emp}
\end{equation}
And with input $\ket{1}\otimes \ket{\phi}\otimes \ket{\psi}$, we get 
\begin{equation}\label{circ:2sameCntrl2b2}
\begin{emp}(50,50)
  setunit 2mm;
      qubits(2);
      gate(gpos 1, btex $B_1$ etex, 0, btex $C_1$ etex);
      cnot(icnd 1, gpos 0);
      widegate(2.5);
      gate(gpos 1, btex $B_3X^{1-l}B_2$ etex, 0, btex $C_2$ etex);
      normalgate;
      cnot(icnd 1, gpos 0);
      gate(gpos 1, btex $B_4$ etex, 0, btex $C_3$ etex);
      label(btex $= $ etex,(QCxcoord+1/2QCstepsize, QCycoord[0]+3mm));
      QCxcoord := QCxcoord + QCstepsize;
      QCstepsize := QCstepsize-4mm;
      wires(2mm);
      gate(icnd 1, gpos 0, btex $X$ etex);
      wires(2mm);
\end{emp}
\end{equation}
Combining circuits~(\ref{circ:2sameCntrl2b1}) and~(\ref{circ:2sameCntrl2b2}), 
we have,
\begin{equation}\label{circ:2sameCntrl2b3}
\begin{emp}(50,50)
  setunit 2mm;
      qubits(2);
      gate(gpos 1, btex $B_1$ etex, 0, btex $C_1$ etex);
      cnot(icnd 1, gpos 0);
      widegate(2); 
      gate(gpos 1, btex $B_2^{\dagger}XB_2$ etex);
      normalgate;
      cnot(icnd 1, gpos 0)
      gate(gpos 1, btex $B_1^{\dagger}$ etex, 0, btex $C_1^{\dagger}$ etex);
      label(btex $= $ etex,(QCxcoord+1/2QCstepsize, QCycoord[0]+3mm));
      QCxcoord := QCxcoord + QCstepsize;
      QCstepsize := QCstepsize-4mm;
      wires(2mm);
      gate(icnd 1, gpos 0, btex $Y$ etex);
      wires(2mm);
\end{emp}
\end{equation}
Similar to the proof for circuit~(\ref{circ:2sameCntrl1b4}), we again 
arrive at a contradiction after considering the trace.~\qed

\medskip

\noindent\textit{Case 3.} 
We now examine five of the remaining six configurations. 

\begin{lemma}
Circuit configurations of the form 
\begin{center}
\begin{emp}(50,50)
  setunit 0.5mm;
  qubits(3);
  conff(1,0,0);
  conff(0,0,1);
  conff(1,1,0);
  conff(0,1,1);
\end{emp}
\end{center}
cannot implement the simplified Toffoli gate~$M$. 
\end{lemma}
\proof 
The most general circuit corresponding to the first configuration in the statement of the lemma is of the form 
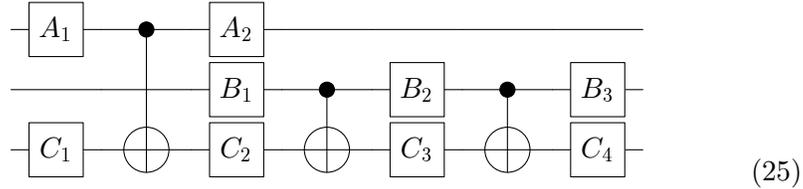
\begin{equation}\label{circ:3cntrla}
\begin{emp}(50,50)
  setunit 2mm;
      qubits(3);
      gate(gpos 2, btex $A_1$ etex, 0, btex $C_1$ etex);
      cnot(icnd 2, gpos 0)
      gate(gpos 2, btex $A_2$ etex, 1, btex $B_1$ etex, 0, btex $C_2$ etex);
      cnot(icnd 1, gpos 0)
      gate(gpos 1, btex $B_2$ etex, 0, btex $C_3$ etex);
      cnot(icnd 1, gpos 0)
      gate(gpos 1, btex $B_3$ etex, 0, btex $C_4$ etex);
\end{emp}
\end{equation}
Following the same steps as in the proofs of 
Lemmas~\ref{lemma:a1sparse} and~\ref{lemma:mprime},
it is straightforward to see that such a circuit cannot realize 
$M$ nor $M'$. 
Therefore,
the first and the third configuration can be ruled out. The other two
configurations can be ruled out by realizing that $M$ is self-inverse.~\qed

\begin{lemma}
The circuit configuration 
\begin{center}
\begin{emp}(50,50)
  setunit 0.5mm;
  qubits(3);
  conff(1,0,1);
\end{emp}
\end{center}
cannot implement the simplified Toffoli gate~$M$. 
\end{lemma}
\proof It suffices to prove that the circuit
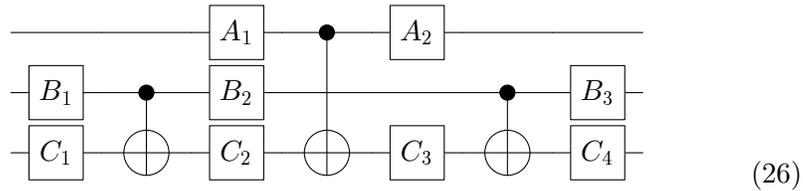
\begin{equation}\label{circ:3cntrlb}
\begin{emp}(50,50)
  setunit 2mm;
      qubits(3);
      gate(gpos 1, btex $B_1$ etex, 0, btex $C_1$ etex);
      cnot(icnd 1, gpos 0)
      gate(gpos 2, btex $A_1$ etex, 1, btex $B_2$ etex, 0, btex $C_2$ etex);
      cnot(icnd 2, gpos 0)
      gate(gpos 2, btex $A_2$ etex, 0, btex $C_3$ etex);
      cnot(icnd 1, gpos 0)
      gate(gpos 1, btex $B_3$ etex, 0, btex $C_4$ etex);
\end{emp}
\end{equation}
cannot realize $M'$. As in Lemma~\ref{lemma:a1sparse}, we note that both $A_1$ and $A_2$ have to be sparse. 
Considering input states of the form $\ket{0}\otimes \ket{\phi}\otimes \ket{\psi}$, it follows from circuit~(\ref{circ:3cntrlb}) that the circuit identity
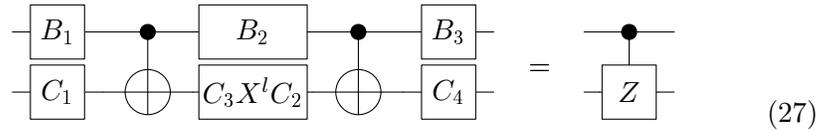
\begin{equation}\label{circ:3cntrlb1}
\begin{emp}(50,50)
  setunit 2mm;
      qubits(2);
      gate(gpos 1, btex $B_1$ etex, 0, btex $C_1$ etex);
      cnot(icnd 1, gpos 0);
      widegate(2);
      gate(gpos 1, btex $B_2$ etex, 0, btex $C_3X^lC_2$ etex);
      normalgate;
      cnot(icnd 1, gpos 0)
      gate(gpos 1, btex $B_3$ etex, 0, btex $C_4$ etex);
      label(btex $= $ etex,(QCxcoord+1/2QCstepsize, QCycoord[0]+3mm));
      QCxcoord := QCxcoord + QCstepsize;
      QCstepsize := QCstepsize-4mm;
      wires(2mm);
      gate(icnd 1, gpos 0, btex $Z$ etex);
      wires(2mm);
\end{emp}
\end{equation}
must hold. We derive from the action on the input 
$\ket{1}\otimes \ket{\phi}\otimes \ket{\psi}$ the circuit identity
\begin{equation}\label{circ:3cntrlb2}
\begin{emp}(50,50)
  setunit 2mm;
      qubits(2);
      gate(gpos 1, btex $B_1$ etex, 0, btex $C_1$ etex);
      cnot(icnd 1, gpos 0);
      widegate(2.5);
      gate(gpos 1, btex $B_2$ etex, 0, btex $C_3X^{1-l}C_2$ etex);
      normalgate;
      cnot(icnd 1, gpos 0);
      gate(gpos 1, btex $B_3$ etex, 0, btex $C_4$ etex);
      label(btex $= $ etex,(QCxcoord+1/2QCstepsize, QCycoord[0]+3mm));
      QCxcoord := QCxcoord + QCstepsize;
      QCstepsize := QCstepsize-4mm;
      wires(2mm);
      gate(icnd 1, gpos 0, btex $X$ etex);
      wires(2mm);
\end{emp}
\end{equation}
Combining circuits~(\ref{circ:3cntrlb1}) and~(\ref{circ:3cntrlb2}),
we have,
\begin{equation}\label{circ:3cntrlb3}
\begin{emp}(50,50)
  setunit 2mm;
      qubits(2);
      gate(gpos 1, btex $B_1$ etex, 0, btex $C_1$ etex);
      cnot(icnd 1, gpos 0);
      widegate(2);
      gate(gpos 0, btex $C_2^{\dagger}XC_2$ etex);
      normalgate;
      cnot(icnd 1, gpos 0)
      gate(gpos 1, btex $B_1^{\dagger}$ etex, 0, btex $C_1^{\dagger}$ etex);
      label(btex $= $ etex,(QCxcoord+1/2QCstepsize, QCycoord[0]+3mm));
      QCxcoord := QCxcoord + QCstepsize;
      QCstepsize := QCstepsize-4mm;
      wires(2mm);
      gate(icnd 1, gpos 0, btex $Y$ etex);
      wires(2mm);
\end{emp}
\end{equation}
The right hand side of~(\ref{circ:3cntrlb3}) cannot produce entangled output states when provided with the input states 
$B_1^{\dagger}\ket{0}\otimes \ket{\phi}$ and 
$B_1^{\dagger}\ket{1}\otimes \ket{\phi}$. 
It follows that either 
\begin{equation}
C_1^{\dagger}C_2^{\dagger}XC_2C_1 = I
\end{equation}
or
\begin{equation}
C_1^{\dagger}XC_2^{\dagger}XC_2XC_1 = I
\end{equation}
holds. Taking the trace on both side, we arrive at a contradiction in
either case.~\qed

\medskip

\noindent\textit{Final Step.}  We have now excluded 26 of the 27
possible control-not configurations. We know that the only viable 
configuration
\begin{center}
\begin{emp}(50,50)
  setunit 0.5mm;
  qubits(3);
  conff(0,1,0);
\end{emp}
\end{center}
actually allows to realize the simplified Toffoli gate~$M$. The
general circuit associated with the circuit configuration is of the
form~(\ref{circ:3cntrlb}). 
\begin{lemma}
At least four single qubit gates are necessary in any realization of
the simplified Toffoli gate $M$ with three controlled-not gates.
\end{lemma}
\proof The only viable general circuit structure with three
controlled-not gate is given by the circuit~(\ref{circ:3cntrlb}). We
will show that at least four single qubit gates in this circuit differ
from multiples of the identity, and that the gate count cannot be
reduced by flipping the gates using~(\ref{circ:flip}).

A circuit~(\ref{circ:3cntrlb}) realizing $M$ is supposed to leave an
input state of the form $\ket{0}\otimes \ket{\varphi}$ invariant. This
implies the circuit identity
\begin{equation}\label{circ:3cntrl4single1}
\begin{emp}(50,50)
  setunit 2mm;
      qubits(2);
      gate(gpos 1, btex $B_1$ etex, 0, btex $C_1$ etex);
      cnot(icnd 1, gpos 0);
      widegate(2);
      gate(gpos 1, btex $B_2$ etex, 0, btex $C_3X^lC_2$ etex);
      normalgate;
      cnot(icnd 1, gpos 0)
      gate(gpos 1, btex $B_3$ etex, 0, btex $C_4$ etex);
      label(btex $= I$ etex,(QCxcoord+1/2QCstepsize, QCycoord[0]+3mm));
      QCxcoord := QCxcoord + QCstepsize;
      QCstepsize := QCstepsize-4mm;
\end{emp}
\end{equation}
And we obtain from the action on $\ket{1}\otimes \ket{\varphi}$ the identity
\begin{equation}\label{circ:3cntrl4single2}
\begin{emp}(50,50)
  setunit 2mm;
      qubits(2);
      gate(gpos 1, btex $B_1$ etex, 0, btex $C_1$ etex);
      cnot(icnd 1, gpos 0);
      widegate(2.5);
      gate(gpos 1, btex $B_2$ etex, 0, btex $C_3X^{1-l}C_2$ etex);
      normalgate;
      cnot(icnd 1, gpos 0)
      gate(gpos 1, btex $B_3$ etex, 0, btex $C_4$ etex);
      label(btex $= $ etex,(QCxcoord+1/2QCstepsize, QCycoord[0]+3mm));
      QCxcoord := QCxcoord + QCstepsize;
      QCstepsize := QCstepsize-4mm;
      wires(2mm);
      gate(icnd 1, gpos 0, btex $Y$ etex);
      gate(gpos 0, btex $Z$ etex);      
      wires(2mm);
\end{emp}
\end{equation}
Combining circuits (\ref{circ:3cntrl4single1}) and (\ref{circ:3cntrl4single2}),
we obtain the equality
\begin{equation}\label{circ:3cntrl4single3}
\begin{emp}(50,50)
  setunit 2mm;
      qubits(2);
      gate(gpos 1, btex $B_1$ etex, 0, btex $C_1$ etex);
      cnot(icnd 1, gpos 0);
      widegate(2);
      gate(gpos 0, btex $C_2^{\dagger}XC_2$ etex);
      normalgate;
      cnot(icnd 1, gpos 0);
      gate(gpos 1, btex $B_1^{\dagger}$ etex, 0, btex $C_1^{\dagger}$ etex);
      label(btex $= $ etex,(QCxcoord+1/2QCstepsize, QCycoord[0]+3mm));
      QCxcoord := QCxcoord + QCstepsize;
      QCstepsize := QCstepsize-4mm;
      wires(2mm);
      gate(icnd 1, gpos 0, btex $Y$ etex);
      gate(gpos 0, btex $Z$ etex);      
      wires(2mm);
\end{emp}
\end{equation}
It follows from Lemma~\ref{lemma:notsparse} that $B_1$
has to be sparse. By Lemma~\ref{lemma:sparse}, this implies that the gates~$B_2$ and $B_3$ in circuit~(\ref{circ:3cntrl4single2}) have to be sparse as well. 
Since we know that $A_1$ and $A_2$ are sparse, flipping 
any number of controlled-not gates in circuit~(\ref{circ:3cntrlb}) 
using equality (\ref{circ:flip}) cannot decrease the count of the 
single-qubit gates. 
It remains to show that none of the four gates $C_1, \dots, C_4$ in 
circuit~(\ref{circ:3cntrlb}) can be a multiple of the identity. 

Since $B_1$ is sparse, let $B_1 = \diag(e^{i\phi_0},e^{i\phi_1})X^k$. 
Considering the input state $\ket{0}\otimes \ket{\phi}$ and 
$\ket{1}\otimes \ket{\phi}$, we can deduce from circuit~(\ref{circ:3cntrl4single3}) the equalities
\begin{eqnarray}
C_1^{\dagger}X^{k}C_2^{\dagger}XC_2X^{k}C_1 = Z, \label{eqn:cc1} \\
C_1^{\dagger}X^{1-k}C_2^{\dagger}XC_2X^{1-k}C_1 = X. \label{eqn:cc2}
\end{eqnarray}
If follows from equations (\ref{eqn:cc1}) and (\ref{eqn:cc2}) that
neither $C_1$ nor $C_2$ can be a multiple of the identity.  Since
$M=M^\dagger$, we can apply the same argument to the inverse of the
circuit~(\ref{circ:3cntrlb}), hence neither $C_3$ nor $C_4$ is a
multiple of the identity either. Therefore, at least four single qubit
gates are non-trivial, as claimed.~\qed

\section{Conclusions}
We have demonstrated that the simplified Toffoli gate by Margolus
cannot be realized with fewer than three controlled-not gates.  Four
additional single qubit gates are required when $M$ is realized with
minimal number of controlled-not gates. Our proof of this lower bound
revealed the interesting fact that the solution by Margolus is
essentially uniquely determined by these constraints.  The tedium of
cases in lower bound proofs can be daunting, but one usually gains
valuable structural insights, particularly if the bounds are tight.
It would be interesting to know tight lower bounds for other
fundamental constructions of quantum circuits, such as the Toffoli
gate. It is conjectured that the Toffoli gate cannot be implemented
with less than six controlled-not and eight single qubit gates.
\medskip

\paragraph{Acknowledgments.}
The research by A.K. was supported by NSF grant EIA 0218582 and a
Texas A\&M TITF initiative.

\end{empfile}

\end{document}